\documentclass[a4paper,11pt]{article}
\pdfoutput=1 

\usepackage{jheppub} 

\usepackage[T1]{fontenc} 
\usepackage{graphicx}
\usepackage{epsfig}  
\usepackage{epsf}    
\usepackage{dcolumn}
\usepackage{bm}
\usepackage{dcolumn}
\usepackage{textcomp}
\usepackage{float}
\usepackage{subfig}
\usepackage[]{hyperref}
  \hypersetup{
  unicode=false,          
  pdftoolbar=true,        
  pdfmenubar=true,        
  pdffitwindow=true,     
  pdfstartview={FitH},    
  pdfsubject={Getting the best out of T2K and NOvA},   
  pdfnewwindow=true,      
  pdfcreator={RevTeX},
  colorlinks=true,       
  linkcolor=red,          
  citecolor=blue,        
  urlcolor=blue,           
  }
\usepackage{hypcap}

\newcommand{\beq}{\begin{equation}}
\newcommand{\eeq}{\end{equation}}
\newcommand{\beqa}{\begin{eqnarray}}
\newcommand{\eeqa}{\end{eqnarray}}

\newcommand{\tx}{{\theta_{12}}}
\newcommand{\ty}{{\theta_{13}}}
\newcommand{\tz}{{\theta_{23}}}

\newcommand{\dl}{{\Delta_{31}}}
\newcommand{\ds}{{\Delta_{21}}}

\newcommand{\atil}{\hat{A}}
\newcommand{\dtil}{\hat{\Delta}}

\newcommand{\dcp}{\delta_{\mathrm{CP}}}
\newcommand{\nova}{NO$\nu$A~}
\newcommand{\pmue}{P(\nu_\mu \rightarrow \nu_e)}
\newcommand{\pme}{P_{\mu e}}

\newcommand{\pmebar}{P_{\bar{\mu} \bar{e}}}

\title{Understanding the degeneracies in \nova data}
\author[a]{Suman Bharti}
\author[b]{Suprabh Prakash}
\author[a]{Ushak Rahaman}
\author[a]{S. Uma Sankar}
\affiliation[a]{Department of Physics, Indian Institute of Technology Bombay,
Mumbai 400076, India}
\affiliation[b]{Instituto de F\'isica Gleb Wataghin - UNICAMP, 13083-859, 
Campinas, SP, Brazil}
\emailAdd{sbharti@phy.iitb.ac.in}
\emailAdd{sprakash@ifi.unicamp.br}
\emailAdd{ushak@phy.iitb.ac.in}
\emailAdd{uma@phy.iitb.ac.in}
\abstract{
The combined analysis of $\nu_\mu$ disappearance and $\nu_e$ appearance data of 
\nova experiment leads to three nearly degenerate solutions. This degeneracy can be understood
in terms of deviations in $\nu_e$ appearance signal, caused by unknown effects, with respect to the signal expected
for a reference set of oscillations parameters.
We define the reference set to be vacuum oscillations in the limit of 
maximal $\tz$ and no CP-violation. We then calculate the 
deviations induced in the $\nu_e$ appearance signal event rate by three unknown effects: (a) matter effects, due to 
normal or inverted hierarchy (b) octant effects, due to $\tz$ being in higher or lower octant and (c)  
CP-violation, whether $\dcp \sim - \pi/2$ or $\dcp \sim \pi/2$. We find
that the deviation caused by each of these effects is the same for
NO$\nu$A. The observed number of $\nu_e$ events in \nova is equivalent to the increase caused by one of the effects.
Therefore, the observed number of $\nu_e$ appearance events of \nova is the net result of  
the increase caused by two of the unknown effects and the decrease caused 
by the third. Thus we get the three degenerate solutions. We also find that further data by \nova can not
distinguish between these degenerate solutions but addition of one
year of neutrino run of DUNE can make a distinction between all three
solutions. The distinction between the two NH solutions and the IH solution
becomes possible because of the larger matter effect in DUNE.
The distinction between the two NH solutions with different octants
is a result of the synergy between the anti-neutrino data of \nova
and the neutrino data of DUNE.}

\begin{document}
\maketitle
\flushbottom

\section{Introduction}
The data from the solar \cite{Bahcall:2004ut, Ahmad:2002jz} and the atmospheric \cite{Ashie:2004mr, Wendell:2010md} neutrino experiments led to the discovery of neutrino oscillations. 
Both the solar and the atmospheric neutrino anomalies can be explained in terms of the oscillations of the three neutrino flavours, $\nu_e$, $\nu_\mu$ and $\nu_\tau$, into one another. 
The oscillation probabilities depend on two independent mass-squared differences, $\ds$ and $\dl$, three mixing angles, $\tx$, $\ty$ and $\tz$, and a CP-violating phase $\dcp$. 
Among these parameters, there are two small quantities: the angle $\ty$ and the ratio $\ds/\dl$. 

During the past decade and a half, a number of experiments with man-made neutrino sources have made precision measurements of the mass-squared differences and the mixing angles. 
This was possible because the expressions for the three flavour survival probabilities reduce to those of effective two flavour survival probabilities, under appropriate approximations. 
For example, setting $\ty=0$ in $P(\bar{\nu}_e\rightarrow\bar{\nu}_e)$ expression for KamLAND \cite{Araki:2004mb,Abe:2008aa} experiment reduces it to an effective two flavour survival probability 
in terms of $\ds$ and $\tx$. A similar effective two flavour survival probability, in terms of $\dl$ and $\tz$, for MINOS \cite{Nichol:2012} experiment can be obtained by setting $\ty=0=\ds$ in the 
expression for $P(\nu_\mu\rightarrow\nu_\mu)$. For the short baseline reactor neutrino experiments, Double-CHOOZ \cite{An:2012eh}, Daya-Bay \cite{Ahn:2012nd} and RENO \cite{Abe:2012tg}, 
an effective two flavour expression in terms of $\dl$ and $\ty$ is obtained by setting $\ds=0$ in the expression for $P(\bar{\nu}_e\rightarrow\bar{\nu}_e)$. This reduction to effective two flavour 
expressions leads to accurate measurement of the modulus of the mass-squared differences and $\sin^22\theta_{ij}$. 
The solar neutrino data requires $\ds$ to be positive but the sign of $\dl$ is still unknown. The value of $\sin^22\theta_{23}$ is measured quite accurately but, since it is close to $1$,
there is a large uncertainty in
the value of $\sin^2\tz$. There is no measurement yet of the CP-violating phase $\dcp$. 
The best-fit values and the allowed $1 \, \sigma$ and $3 \, \sigma$ 
of the mass-squared differences and the mixing angles from the disappearance data of the above experiments plus the solar and the atmospheric data is given in table~\ref{globalneutrinodatatable}.

\begin{table}[htbp]
\begin{center}
 \begin{tabular}{|l|l|l|l|l|}
\hline \hline
    Parameter                                                        & Best fit   & 1$\sigma\ {\rm range}$             & 3$\sigma\ {\rm range}$   \\ 
\hline \hline
   $\delta m^2/10^{-5} {\rm\ eV^2\ (NH\ or\ IH)}$                    & 7.50       & 7.33 - 7.69                        & 7.03 - 8.09              \\ 
\hline
    $\sin^2\theta_{12} {\rm\ (NH\ or\ IH)}$                          & 0.306      & 0.294 - 0.318                      & 0.271 - 0.345            \\
\hline
   $\Delta m^2/10^{-3} {\rm\ eV^2\ (NH)}$                            & 2.524      & 2.484 - 2.563                      & 2.407 - 2.643            \\
   $\Delta m^2/10^{-3} {\rm\ eV^2\ (IH)}$                            & -2.514     & -2.555 - -2.476                    & -2.635 - -2.399          \\
\hline
   $\sin^2\theta_{13} {\rm\ (NH)}$                                   & 0.02166    & 0.02091 - 0.02241                  & 0.01934 - 0.02392        \\
   $\sin^2\theta_{13} {\rm\ (IH)}$                                   & 0.02179    & 0.02103 - 0.02255                  & 0.01953 - 0.02408        \\
\hline
   $\sin^2\theta_{23} {\rm\ (NH)}$                                   & 0.441      & 0.420 - 0.468                      & 0.385 - 0.635            \\
   $\sin^2\theta_{23} {\rm\ (IH)}$                                   & 0.587      & 0.563 - 0.607                      & 0.393 - 0.640            \\
\hline \hline  
  \end{tabular}

\caption{\footnotesize{Neutrino mass-squared differences and mixing angles from global analysis of solar, atmospheric, reactor and accelerator data \cite{Esteban:2016qun}.
Note that \nova data is not included in this analysis.}}
\label{globalneutrinodatatable}
\end{center}
\end{table}

At present the two long baseline accelerator experiments, T2K and NO$\nu$A, are taking data \cite{Abe:2011ks, T2Kapp, T2Kdisapp, nova_tdr}. These experiments observe $\nu_\mu\rightarrow\nu_e$ 
appearance as well as $\nu_\mu$ disappearance. The dominant oscillations for these experiments are driven by $\dl$. These experiments are also designed to be sensitive to 
CP-violation in neutrino oscillations. Hence they are also sensitive to $\ds$ dependent sub-dominant term in the oscillation probability. Thus the data of these two 
experiments must necessarily be analysed using the full three flavour expressions for the neutrino survival ($\nu_\mu$ disappearance) and oscillation ($\nu_e$ appearance) 
probabilities. Since these probabilities depend on a number of parameters, degenerate solutions arise when they are fit to the data. In particular, the $\nu_\mu\rightarrow\nu_e$ 
appearance probability depends on three unknowns: (a) neutrino mass hierarchy ($\dl > 0$ or $\dl < 0$), (b) $\tz$ octant ($\tz >\pi/4$ or $\tz < \pi/4$) and (c) value of $\dcp$. In this report, we study how the three degenerate 
solutions of \nova arise due to the above three unknowns. We also investigate how the DUNE \cite{Abi:2017aow} experiment can fully resolve this three fold degeneracy.     

\section{Degeneracies in $\pmue$ }

In T2K and \nova experiments, the neutrinos travel long distances through
earth matter and undergo coherent forward scattering. The effect of this scattering
is taken into account through the Wolfenstein matter term \cite{msw1}
\begin{equation}
A \ ({\rm in \ eV^2}) = 0.76 \times 10^{-4} \rho \ ({\rm in \ gm/cc}) \ E \ ({\rm in \ GeV}),
\end{equation}
where $E$ is the energy of the neutrino and $\rho$ is the density of the matter.
The interference between $A$ and $\Delta_{31}$ leads to the modification of neutrino
oscillation probability due to matter effects. This modified expression for $\pmue$
is given by \cite{Cervera:2000kp,Freund:2001pn} 
\begin{eqnarray}
\pmue & = & \pme =
\sin^2 2 \ty \sin^2 \tz\frac{\sin^2\dtil(1-\atil)}{(1-\atil)^2} \nonumber\\ 
& & +\alpha \cos \ty \sin2\tx \sin 2\ty \sin 2\tz \cos(\dtil+\dcp)
\frac{\sin\dtil \atil}{\atil} \frac{\sin \dtil(1-\atil)}{1-\atil} \nonumber\\
 & & +\alpha^2 \sin^2 2 \tx \cos^2 \ty \cos^2 \tz 
 \frac{\sin^2 \dtil \atil}{\atil^2},
\label{pmue-exp}
\end{eqnarray}
where $\hat{\Delta} = 1.27 \Delta_{31}L/E$, $\hat{A} = A/\Delta_{31}$ and 
$\alpha = \Delta_{21}/\Delta_{31}$. For anti-neutrinos, 
$P(\bar{\nu}_\mu \to \bar{\nu}_e) = \pmebar$ is given by a similar expression
with $\dcp \to - \dcp$ and $A \to -A$. Since $\alpha\approx 0.03$, the term proportional to $\alpha^2$ in $\pme$ can be neglected. Since the experiments are designed to be sensitive to $\dcp$, 
the second term, proportional to $\alpha$ must be retained. If $\dcp$ is in the lower half plane (LHP, $-180^\circ\leq\dcp\leq 0$) $\pme$ is larger compared to the CP conserving case whereas it is smaller for $\dcp$
in the upper half plane (UHP, $0\leq\dcp\leq 180^\circ$). For $\pmebar$ the situation is reversed. For the purpose of discussion in the paragraph below, we take $\dcp$
to be a binary variable which either increases $\pme$ or decreases it.

The dominant term in $\pme$ is proportional to $\sin^2 2\ty$ and hence this probability is rather small. Matter effect can enhance (suppress) it by about $22\%$ for \nova if $\dl$ is 
positive (negative) \cite{Narayan:1999ck}. The situation is opposite for $\pmebar$.
This dominant term is also proportional to $\sin^2\tz$. For $\sin^2 2 \tz < 1$,
	there are two possible solutions: One with $\sin^2 \tz > 0.5$ and the other with $\sin^2 \tz < 0.5$.
	In the former (latter) case, $\pme$ is enhanced (suppressed) relative to the case of 
	maximal mixing.  
Since each of the unknowns can take two possible values, there are eight different combinations 
of the three unknowns. A given value of $\pme$ can be reproduced, for any combination of the unknowns, by 
choosing the value of $\ty$ appropriately. Thus there is an eight-fold degeneracy in interpreting the expression for $\pme$, if the value of $\sin^2 2\ty$ is not known precisely. The degeneracy between $\ty$ and non-maximal 
$\tz$ was pointed out in \cite{Fogli:1996pv} whereas the degeneracy between $\ty$ and $\dcp$ in
$\pme$ was highlighted in \cite{BurguetCastell:2001ez}. The degeneracy between the sign of $\Delta_{31}$
and $\dcp$ was studied in \cite{Minakata:2001qm,Mena:2004sa,Prakash:2012az} and that between non-maximal
$\tz$ and $\dcp$ was considered in \cite{Meloni:2008bd,Agarwalla:2013ju,Nath:2015kjg,Bora:2016tmb}. 
Possible methods to resolve the eight-fold degeneracy were discussed in \cite{Barger:2001yr,Kajita:2006bt}.
 We will show below that the present precision measurement of $\ty$ breaks this eight-fold degeneracy into $(1+3+3+1)$ pattern.    

\section{$\nu_e$ appearance events in \nova}

\subsection{2017 analysis}
\nova \cite{nova_tdr} is a long baseline neutrino oscillation experiment capable
of measuring the survival probability $P(\nu_\mu \to \nu_\mu)$
and the oscillation probability $P(\nu_\mu \to \nu_e)$. The 
NuMI beam at Fermilab, with a power of 700 kW which corresponds
to $6\times 10^{20}$ protons on target (POT) per year, produces the neutrinos. 
The far detector consists of 14 kton of totally active scintillator 
material and is located 810 km away at a $0.8^\circ$ off-axis location.
Due to the off-axis location, the flux peaks sharply at
2 GeV, which is close to the energy of maximum oscillation of 1.4 GeV.
It has started taking data in 2014 and is expected to run three years in
neutrino mode and three years in anti-neutrino mode. The combined analysis of $\nu_\mu$ disappearance and  $\nu_e$ appearance data is given in ref.~\cite{Adamson:2017gxd},
which is based on a neutrino run with $6.05\times 10^{20}$ POT. This analysis gives the following three (almost) degenerate solutions for the unknown quantities: 
\begin{enumerate}
\item normal hierarchy ($\dl$ +ve), $\sin^2\tz=0.4$, $\dcp=-90^\circ$ (NH, LO, $-90^\circ$),
\item normal hierarchy ($\dl$ +ve), $\sin^2\tz=0.62$, $\dcp=135^\circ$ (NH, HO, $135^\circ$) and
\item inverted hierarchy ($\dl$ -ve), $\sin^2\tz=0.62$, $\dcp=-90^\circ$ (IH, HO, $-90^\circ$).
\end{enumerate} 

To understand the existence of the above three solutions, we first calculate $\nu_e$ appearance in \nova for the case of vacuum oscillations with $\tz=45^\circ$ and $\dcp=0$. 
We then consider the changes in this number due to (a) matter effects, (b) $\tz$ octant effect and (c) large value of $\dcp$. First we introduce one change at a time in the following manner:
\begin{itemize}
\item normal hierarchy (NH), which increases $\pme$ or inverted hierarchy (IH) which decreases it,
\item higher octant (HO), which increases $\pme$ or lower octant (LO) which decreases it and
\item $\dcp=-90^\circ$, which increases $\pme$ or $\dcp=+90^\circ$, which decreases it.
\end{itemize}
The event numbers are calculated using GLoBES software \cite{Huber:2004ka,Huber:2007ji}. The following inputs are used for the well-measured neutrino 
parameters: $\ds = 7.5\times 10^{-5}$ eV$^2$, $\sin^2\tx = 0.306$, $\dl{\rm (NH)} = 2.74\times 10^{-3}$ eV$^2$, $\dl{\rm (IH)} = -2.65\times 10^{-3}$ eV$^2$ and $\sin^2 2\ty = 0.085$.
The values of $\dl {\rm (NH)}$ and $\dl {\rm (IH)}$ are taken from the fit to \nova disappearance data \cite{Adamson:2017qqn}. The inputs for the undetermined parameters are taken to be one of three 
possible values: the reference value mentioned at the beginning of the paragraph (labelled `$0$'), the value which increases $\pme$ (labelled `$+$') and the one which decreases it (labelled `$-$').   
The results are displayed in table~\ref{Nu-table-1}. From this table, we note that the increase   in $\nu_e$ appearance events, for any single `+' change of the undetermined parameters, 
is essentially the same. A similar comment applies to the case of any single `-' change.

\begin{table}
\begin{center}
\begin{tabular}{ |c|c|c|c|c| }
\hline\hline   
Hierarchy--$\sin^2\theta_{23}$--$\delta_{cp}$ & Label & Signal eve. & Bg eve. & Total eve. \\
\hline\hline
Vac.--0.5--0 & (0 0 0) & 20.17 & 6.32 & 26.49 \\
\hline
NH--0.5--0 & (+ 0 0) & 24.95 & 6.33 & 31.28 \\
\hline
IH--0.5--0 & (- 0 0) & 14.90 & 6.18 & 21.08 \\
\hline      
Vac.--0.5-- -90 & (0 0 +) & 24.68 & 6.36 & 31.04 \\
\hline   
Vac.--0.5-- +90 & (0 0 -) & 14.82 & 6.36 & 21.18 \\
\hline
Vac.--0.62--0 & (0 + 0) & 24.73 & 8.15 & 32.88 \\ 
 \hline  
Vac.--0.4--0 & (0 - 0) & 16.50 & 8.09 & 24.59 \\
\hline
\hline
\end{tabular}
\caption{\footnotesize{ Number of $\nu_e$ appearance events for one year $\nu$ run of NO$\nu$A. They are listed for the reference point and for change of one unknown at a time.}}
\label{Nu-table-1}
\end{center}
\end{table}

Next we consider all the eight possible combinations in the changes of the three undetermined parameters. For example, all three may shift in such a way that each shift leads to increase in $\pme$. 
We label this case as $(+++)$. In such a case, we get the maximum number of $\nu_e$ appearance events. Another case is that two of the undetermined parameters shift so as to increase $\pme$ whereas 
the third parameter shifts to lower it. This can occur in three possible ways, which we label as $(++-)$, $(+-+)$ and $(-++)$. These three combinations predict a moderate increase in the number of $\nu_e$ 
appearance events compared to the reference case. Similarly there is a case where shift in one parameter increases $\pme$
but shifts in the other two lower it, with the three possibilities $(+--)$, $(-+-)$ and $(--+)$, which predict a moderate decrease in the number of $\nu_e$ 
appearance events compared to the reference case. Finally, there is a case where each of the three shifts lowers $\pme$, labelled $(---)$, which predicts the smallest number of $\nu_e$ appearance events. 
The number of $\nu_e$ appearance events for NO$\nu$A, for each of the above eight combinations, are listed in table~\ref{Nu-table-2}, which are also calculated using GLoBES. From this table, we note
that the number of events for the three combinations $(++-)$, $(+-+)$ and $(-++)$ are nearly the same. Such a statement is also true for $(+--)$, $(-+-)$ and $(--+)$ combinations. 
The predictions for the combinations $(+++)$ and $(---)$ are unique. 
Thus the eight-fold degeneracy, which was present when $\ty$ was not measured, splits into $(1+3+3+1)$ pattern with the precision measurement of $\ty$ \cite{An:2016ses}, as mentioned in the introduction. 
The $\nu_e$ appearance data of \nova shows a modest increase relative to the reference case. Hence there is a three-fold degeneracy in \nova solutions.
The predictions of $\nu_e$ appearance events, for each of the three \nova solutions, are listed in table~\ref{nova-table}. The predictions for the two NH solutions matched the experimental numbers.
The prediction for the IH solution, which is $0.5\,\sigma$ away from the NH solutions, is lower by $3$ (half the statistical uncertainty in the expected number).
A more detailed calculation shows that the agreement is valid for $\nu_e$ appearance spectrum also. We have verified this through GLoBES simulations. The occurance of three fold degenerate solutions of \nova, 
based on the degeneracies inherent in $\pme$, was discussed previously in ref.\cite{Lindner:2017nmb}.

\begin{table}
\begin{center}
\begin{tabular}{ |c|c|c|c|c| }
\hline\hline   
Hierarchy--$\sin^2\theta_{23}$--$\delta_{cp}$ & Label & Signal eve. & Bg eve. & Total eve. \\
\hline\hline
NH--0.62-- -90 & (+ + +) & 35.59 & 8.08 & 43.67 \\
\hline
NH--0.4-- -90 & (+ - +) & 25.34 & 8.20 & 33.54 \\
\hline
NH--0.62-- +90 & (+ + -) &  24.96 & 8.08 & 33.04 \\ 
\hline
IH--0.62-- -90 & (- + +) & 22.80 & 8.14 & 30.94 \\
\hline
NH--0.4-- +90 & (+ - -) & 14.59 & 8.20 & 22.79 \\
\hline     
IH--0.4-- -90 & (- - +) & 16.59 & 7.88 & 24.47 \\
\hline   
IH--0.62-- +90 & (- + -) & 14.49 & 8.14 & 22.63 \\ 
 \hline
IH--0.4-- +90 & (- - -) & 8.19 & 7.88 & 16.07 \\
\hline   
\hline
\end{tabular}
\caption{\footnotesize{ Number of $\nu_e$ appearance events for one year $\nu$ run of NO$\nu$A, for the eight different combinations of unknowns.}} 
\label{Nu-table-2}
\end{center}
\end{table}

\begin{table}
\begin{center}
\begin{tabular}{ |c|c|c|c|c| }
\hline\hline   
Hierarchy--$\sin^2\theta_{23}$--$\delta_{cp}$ & Label & Signal eve. & Bg eve. & Total eve. \\
 \hline\hline
NH--0.404-- -86 & (+ - +) & 25.35 & 8.20 & 33.55 \\
\hline   
NH--0.62-- +135 & (+ + -) &  26.24 & 8.12 & 34.36 \\ 
\hline   
IH--0.62-- -90 & (- + +) & 22.80 & 8.14 & 30.94 \\
\hline   
\hline
\end{tabular}
\caption{\footnotesize{Number of expected $\nu_e$ appearance events for one year $\nu$ run of NO$\nu$A, for the three solutions in ref.~\cite{Adamson:2017gxd}.}} 
\label{nova-table}
\end{center}
\end{table}

We now consider if it is possible to resolve this three-fold degeneracy with anti-neutrino data from NO$\nu$A. 
For anti-neutrinos the sign of matter term $A$ is reversed and 
so is the sign of $\dcp$. The probability $\pmebar$ decreases (increases) for NH (IH). 
It also increases (decreases) for $\dcp$ in UHP (LHP). However, we will continue to label NH by `+' and IH by `-'. Similarly we will label LHP by `+' and UHP by `-'. 
But, it should be remembered that, for anti-neutrinos, `+' sign (`-' sign) for hierarchy and $\dcp$ leads to a decrease (increase)
in event rates. For octant, however, `+' sign (`-' sign) lead to increase (decrease) in $\pmebar$ also. For the $(++-)$ solution of NO$\nu$A the value of $\pmebar$ decreases 
due to hierarchy and increases due to octant and $\dcp$. For the $(-++)$ solution, $\pmebar$ increases due to hierarchy and octant and decreases due to $\dcp$. 
Hence, these two solutions are degenerate for anti-neutrino data also and \nova data can not distinguished between them. 
In the case of the third $(+-+)$ solution, $\pmebar$ decreases due to all three unknowns. The prediction for $\bar{\nu}_e$ appearance 
events for this solution will be the smallest. In principle, the anti-neutrino data of \nova should distinguish the $(+-+)$ solution 
from the other two.  
Since the expected number of $\bar{\nu}_e$ appearance events for this case are particularly small, 
as we can see from 
table~\ref{Nubar-table}, the statistical uncertainties are very large. 
So it is difficult for \nova to distinguish this solution from the other two at $3~\sigma$ level.

\begin{table}
\begin{center}
\begin{tabular}{ |c|c|c|c|c| }
\hline\hline   
Hierarchy--$\sin^2\theta_{23}$--$\delta_{cp}$ & Lable & Signal eve. & Bg eve. & Total eve. \\
 \hline\hline
NH--0.404-- -86 & (+ - +) & 3.04 & 3.81 & 6.85 \\
\hline   
NH--0.62-- +135 & (+ + -) & 7.83 & 3.95 & 11.78 \\
\hline   
IH--0.62-- -90 & (- + +) & 9.09 & 3.77 & 12.86 \\
\hline   
\hline
\end{tabular}
\caption{\footnotesize{Number of expected $\bar{\nu}_e$ appearance events for one year $\bar{\nu}$ run of NO$\nu$A, for the three solutions in ref.~\cite{Adamson:2017gxd}. }}
\label{Nubar-table}
\end{center}
\end{table}

\begin{figure}[t]
\centering
\includegraphics[width=1.0\textwidth]{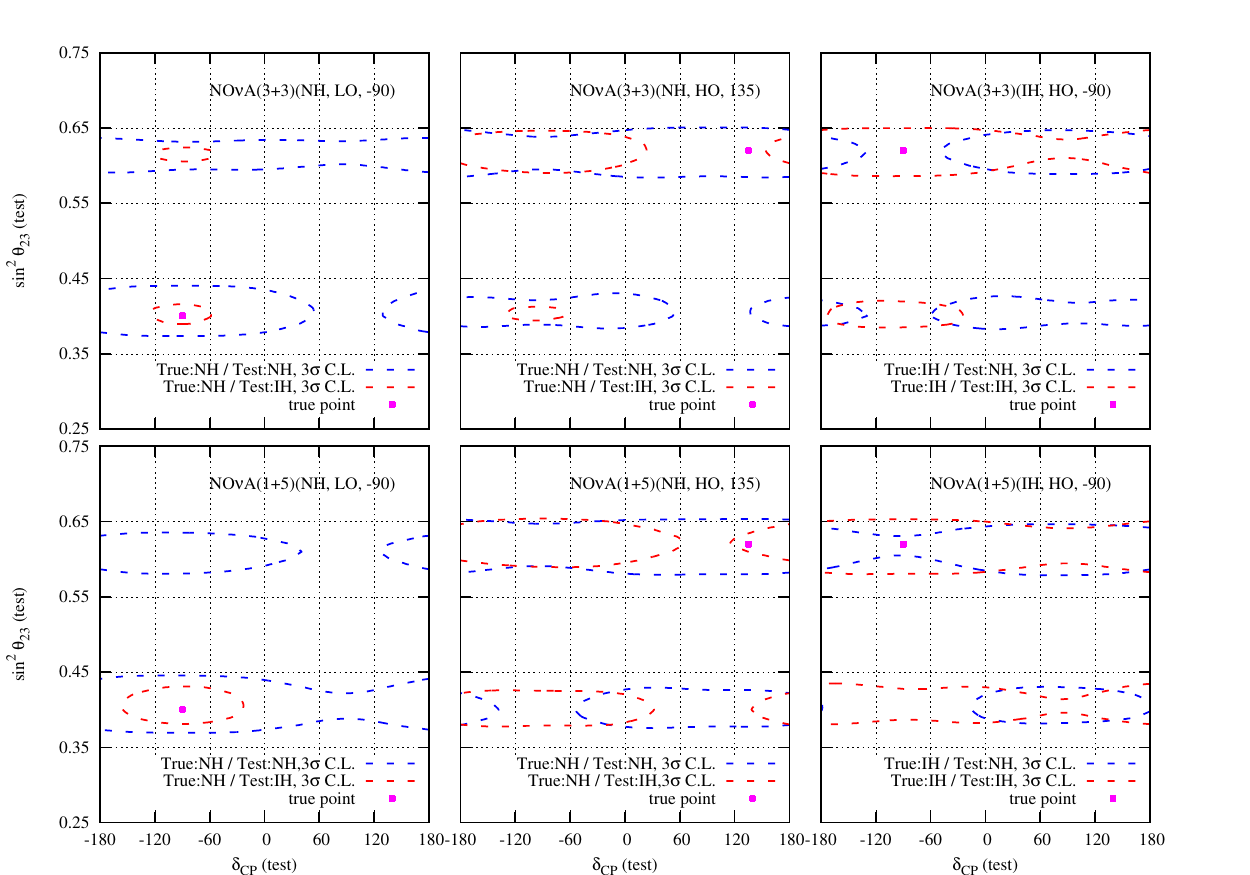}
\caption{\footnotesize{Expected allowed regions in $sin^2\theta_{23}-\delta_{cp}$ plane for a six years run of NO$\nu$A, assuming one of the 
best-fit points is the true solution. The left, middle and right columns represent  (NH, LO, $-90^\circ$), (NH, HO, $135^\circ$) and (IH, HO, $-90^\circ$) solutions respectively. 
The top and bottom rows are for (3$\nu$ + 3$\bar{\nu}$) and (1$\nu$ + 5$\bar{\nu}$)
 runs of \nova respectively. The blue (red) curves are there for test NH (IH).}}
\label{nova-degeneracy}
\end{figure}

We illustrate our results in figure~\ref{nova-degeneracy}. The plots in this 
figure are prepared using the following procedure.
Each of the three solutions was used as the 
input point in GLoBES to obtain disappearance and appearance event spectra of NO$\nu$A for a three year neutrino run 
and a three year anti-neutrino run, which we label as 
(3$\nu$ + 3$\bar{\nu}$) run. The input values of the fixed neutrino parameters are given previously. These spectra are contrasted with 
the simulated spectra, also calculated using GLoBES, 
where the test values of undetermined parameters are varied over the following ranges: test hierarchy -- NH or IH, 
test $\sin^2\tz$ -- $(0.3,0.7)$ and test $\dcp$ -- $(-180^\circ, +180^\circ)$. 
The $\chi^2$ between the spectra with \nova best-fit point as input and the simulated spectra with test values as input is computed. 
The plots in the top row show the allowed regions 
in $\dcp-\sin^2\tz$ plane at $3~\sigma$ C.L., by this (3$\nu$ + 3$\bar{\nu}$) run. We see that for a given solution of \nova the other 
two solutions are not ruled out at $3~\sigma$ level \cite{Goswami:2017hcw}. 
These plots show that the two solutions, (NH, HO, $135^\circ$) and (IH, HO, $-90^\circ$) can not be distinguished from each other, 
as explained above. Due to the anti-neutrino data, 
the (NH, LO, $-90^\circ$) solution is partly isolated. But the large statistical errors in the anti-neutrino data 
do not allow a complete isolation. We have also done the simulation 
for a one year neutrino run followed by a five year anti-neutrino run of NO$\nu$A, which we label as (1$\nu$ + 5$\bar{\nu}$) run. 
It was hoped that the increased anti-neutrino statistics may help in 
isolating the (NH, LO, $-90^\circ$) solution. Despite the increased exposure, the number of $\bar{\nu}_e$ appearance events 
is too small to distinguish the (NH, LO, $-90^\circ$) solution. 
This can be seen from the three plots in the bottom row of 
figure~\ref{nova-degeneracy}. We have also checked the discrimination 
capabilities of (4$\nu$ + 2$\bar{\nu}$) and (2$\nu$ + 4$\bar{\nu}$) runs. They are not noticeably 
different from those of the (3$\nu$ + 3$\bar{\nu}$) run.

\begin{figure}[t]
\centering
\includegraphics[width=1.0\textwidth]{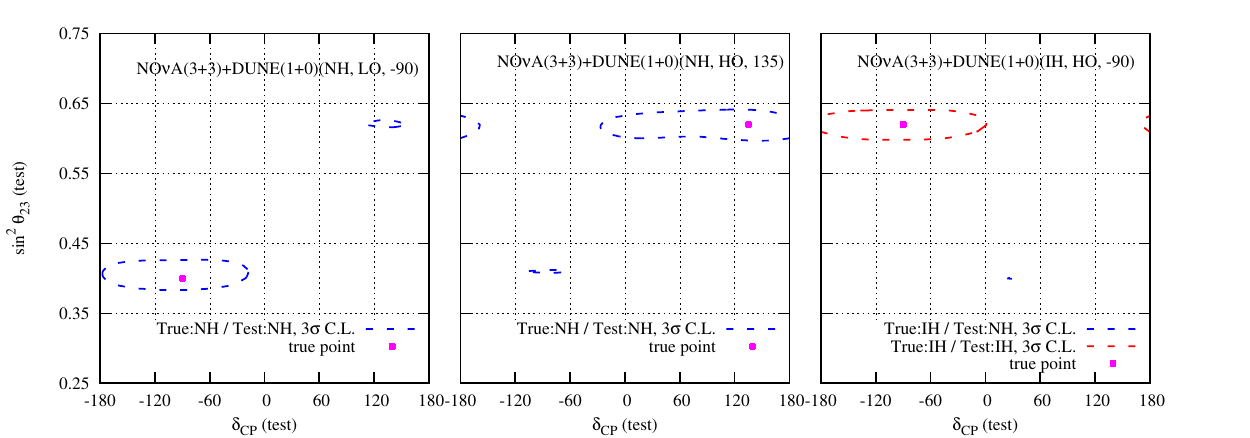}
\caption{\footnotesize {Expected allowed regions in $\sin^2\theta_{23}-\delta_{cp}$ plane for a (3$\nu$ + 3$\bar{\nu}$) run of NO$\nu$A plus a one year neutrino run of DUNE, assuming one of the 
three solutions in ref.~\cite{Adamson:2017gxd} is the true solution. The left, middle and right panels represent (NH, LO, $-90^\circ$), (NH, HO, $135^\circ$) and (IH, HO, $-90^\circ$) solutions respectively. 
The blue (red) curves are there for test NH (IH).}}
\label{nova+dune}
\end{figure}

From figure~\ref{nova+dune}, we see that the addition of one year of neutrino 
data of DUNE to \nova data of (3$\nu$ + 3$\bar{\nu}$) run leads to an 
essentially unique identification
of the correct solution at $3 \sigma$ level. The average neutrino energy for the DUNE experiment is larger than the energy of \nova and hence its 
matter effect is larger. Therefore, the change in $\nu_e$ appearance events induced by matter effects is larger compared to the changes induced by 
octant effects or by $\dcp$. This sets apart the IH solution from the two NH solutions. There is a modest difference in the prediction 
of $\nu_e$ appearance events for the two NH solutions with different octants of $\tz$, as shown in table~\ref{DUNE-Nu-table-1}. This difference, combined with the 
discriminating power of \nova anti-neutrino data, leads to a $3~\sigma$ distinction between the two NH solutions. 
Thus the synergy between the anti-neutrino data of \nova and the neutrino data of DUNE plays an important role in
distinguishing between the two NH solutions. In ref.~\cite{Goswami:2017hcw} the combination of
\nova (3$\nu$ + 3$\bar{\nu}$) run along with DUNE (1$\nu$ + 1$\bar{\nu}$) run was considered. Their results are very similar to our results. We have not included T2K in these simulations because
its best-fit value of $\sin^2\tz$ \cite{T2Kdisapp} does not agree with any of the solutions given in ref.~\cite{Adamson:2017gxd}.

\begin{table}
\begin{center}
\begin{tabular}{ |c|c|c|c| }
\hline\hline   
Hierarchy--$\sin^2\theta_{23}$--$\delta_{cp}$ & Signal eve. & Bg eve. & Total eve. \\
 \hline\hline
NH--0.404-- -86 & 332.22 & 97.91 & 430.13 \\
\hline
NH--0.62-- +135 & 353.12 & 97.81 & 450.93 \\
\hline
IH--0.62-- -90 & 220.09 & 100.03 & 320.12 \\
\hline   
\hline
\end{tabular}
\caption{\footnotesize{Number of expected $\nu_e$ appearance events for one year $\nu$ run of DUNE, for the three solutions in ref.~\cite{Adamson:2017gxd}.}}
\label{DUNE-Nu-table-1}
\end{center}
\end{table}


\subsection{2018 data}
During the past year, the \nova collaboration has re-calibrated their signal identification algorithms \cite{Jan-2018}. In addition they have accumulated more data with a total POT of
$8.85 \times 10^{20}$ \cite{NOvA:2018gge}.
As a result of the analysis with the new procedure, \nova finds a best-fit solution in the  higher octant at (NH, $\sin^2\tz=0.56$, $\dcp=-144^\circ$). There is a nearly 
degenerate solution in the lower octant at (NH, $\sin^2\tz=0.47$, $\dcp=-72^\circ$). There is no IH solution at $1~\sigma$. In this subsection, we discuss the ability of long
baseline neutrino experiments to distinguish between these two solutions.  

We study this discrimination ability using the same procedure as before.
The parameters of the HO solution are used as input to GLoBES and the neutrino and anti-neutrino event spectra are simulated. We also use 
GLoBES to simulate these spectra for various `test' values of the neutrino
oscillation parameters and compute the $\chi^2$ between the spectrum of
the HO solution and each of the test spectra. This computation is done for
for three different situations: for \nova simulations alone, for \nova + T2K simulations and for \nova + T2K + DUNE simulations. The same procedure is repeated for the 
LO solution. In dealing with these new solutions, we have included the simulation of T2K data also, because the newly allowed values of $\sin^2\tz$ agree with T2K best-fit value \cite{T2Kdisapp}.  

\begin{figure}[t]
\centering
\includegraphics[width=1.0\textwidth]{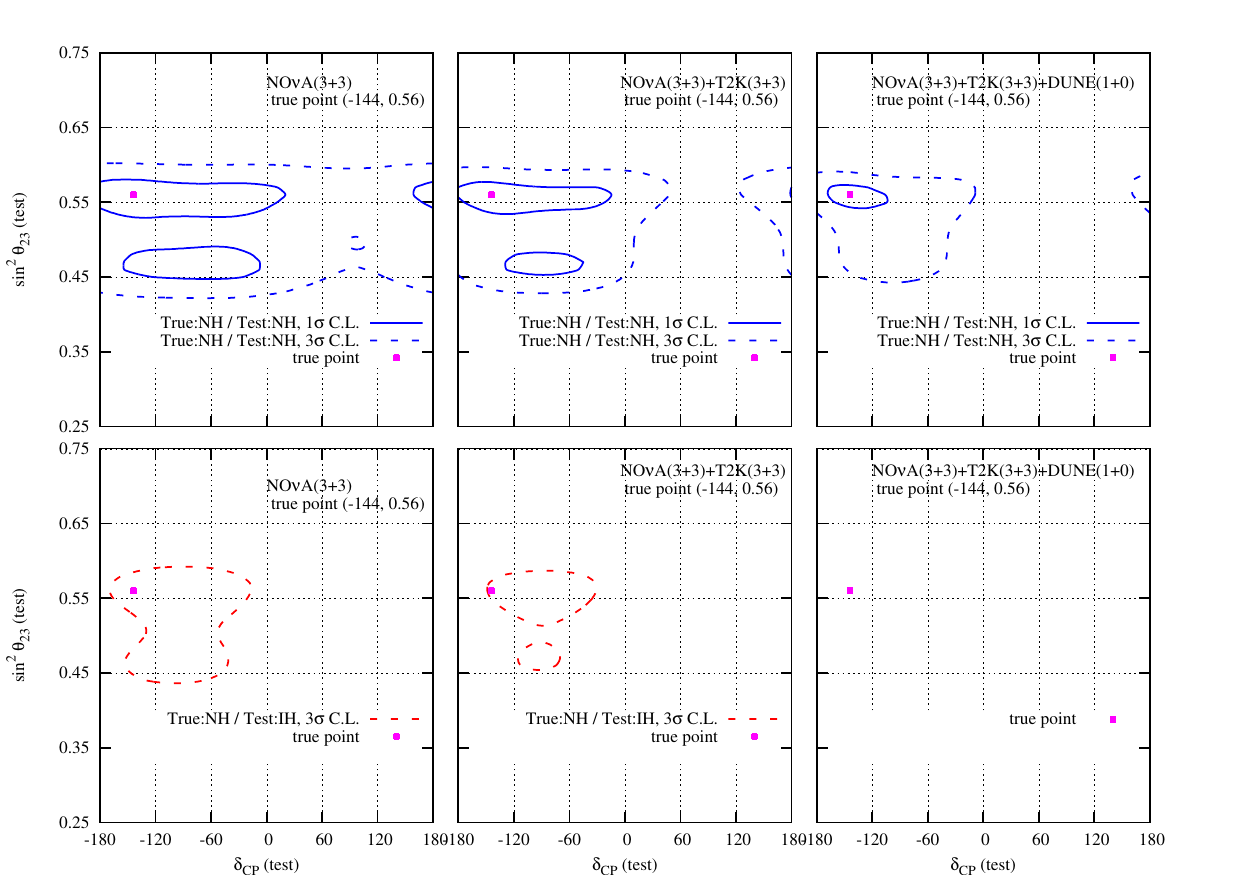}
\caption{\footnotesize{Allowed regions in $sin^2\theta_{23}-\delta_{cp}$ parameter space assuming the HO solution is correct. 
The left, middle and right columns are for NO$\nu$A $(3\nu+3\bar{\nu})$, NO$\nu$A $(3\nu+3\bar{\nu})$ + T2K $(3\nu+3\bar{\nu})$ 
and NO$\nu$A $(3\nu+3\bar{\nu})$ + T2K $(3\nu+3\bar{\nu})$ + DUNE $(1\nu)$  respectively. The top (bottom) row is for test NH (IH).}}
\label{HO2018}
\end{figure}

\begin{figure}[t]
\centering
\includegraphics[width=1.0\textwidth]{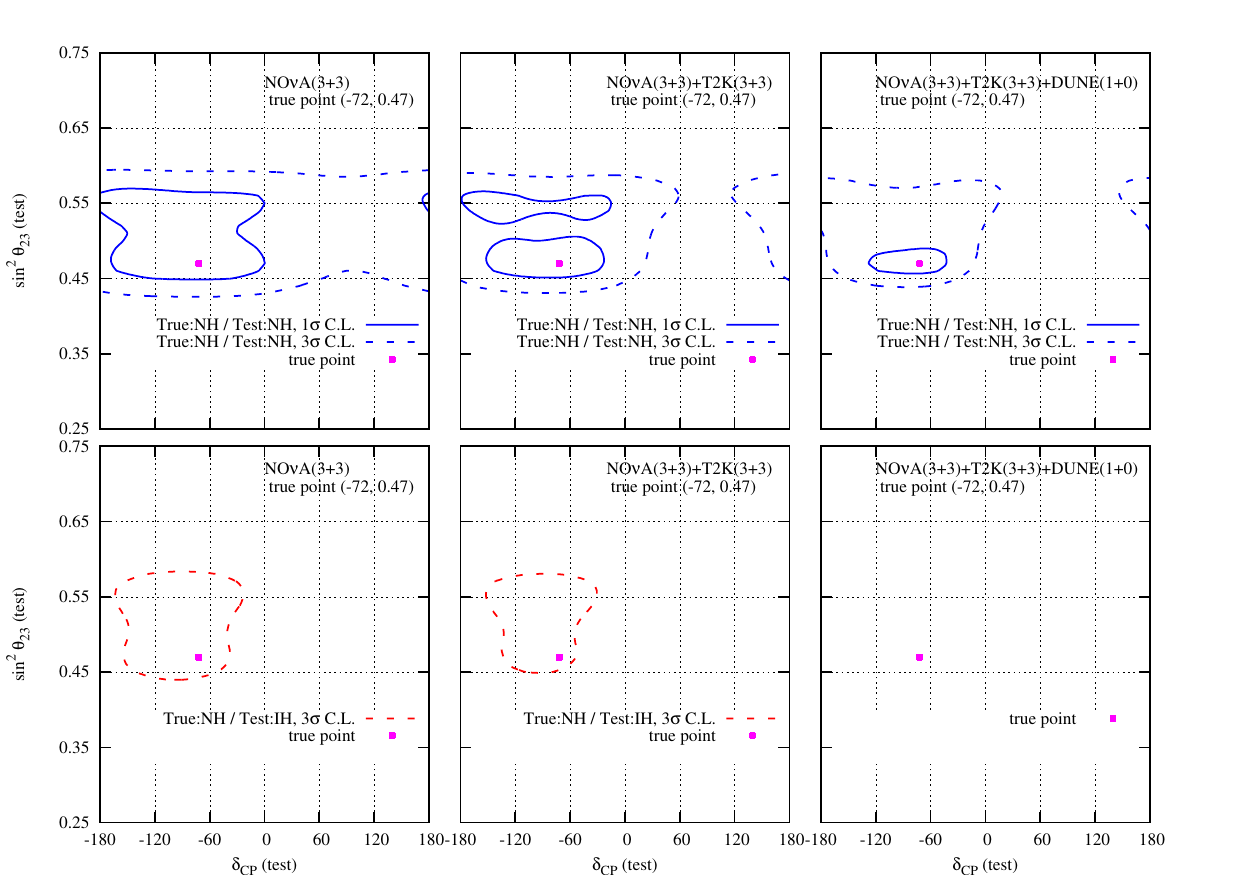}
\caption{\footnotesize{Allowed regions in $\sin^2\theta_{23}-\delta_{cp}$ parameter space assuming the LO solution is correct. 
The left, middle and right columns are for NO$\nu$A $(3\nu+3\bar{\nu})$, NO$\nu$A $(3\nu+3\bar{\nu})$ + T2K $(3\nu+3\bar{\nu})$ 
and NO$\nu$A $(3\nu+3\bar{\nu})$ + T2K $(3\nu+3\bar{\nu})$ + DUNE $(1\nu)$  respectively. The top (bottom) row is for test NH (IH).}}
\label{LO2018}
\end{figure}

The results are shown in figure~\ref{HO2018} and figure~\ref{LO2018}. In each 
figure, NH (IH) is the test hierarchy for the top (bottom) row. The plots
in the left column are for \nova $(3\nu+3\bar{\nu})$ run, those in the middle column are
for the combination of \nova $(3\nu+3\bar{\nu})$ run and T2K $(3\nu+3\bar{\nu})$ run and those in the
right column are for the above combination along with a one year neutrino
run of DUNE. We see from the bottom rows of these two figures that 
neither \nova alone nor \nova + T2K can rule out the IH (the wrong hierarchy) at $3~\sigma$ level. However, the addition of one year of 
neutrino data from DUNE is very effective in ruling out the wrong hierarchy.

Turning our attention to the discrimination between the two different
octant solutions, we find that the data from \nova alone can not distinguish between them. Addition of T2K data helps in reducing the 
allowed regions a little but still does not provide a discrimination
between the two octants. T2K data strongly discriminates against $\dcp
\approx 90^\circ$ hence the test values around this region are ruled out
at $3~\sigma$, though they are allowed by \nova data. One year neutrino data of DUNE, which has a modest octant discrimination power, is able to rule out the wrong octant 
at $1~\sigma$ but not at $3~\sigma$. Neither 
\nova nor \nova + T2K can establish CP violation at $3~\sigma$.

\begin{figure}[t]
\centering
\includegraphics[width=0.8\textwidth]{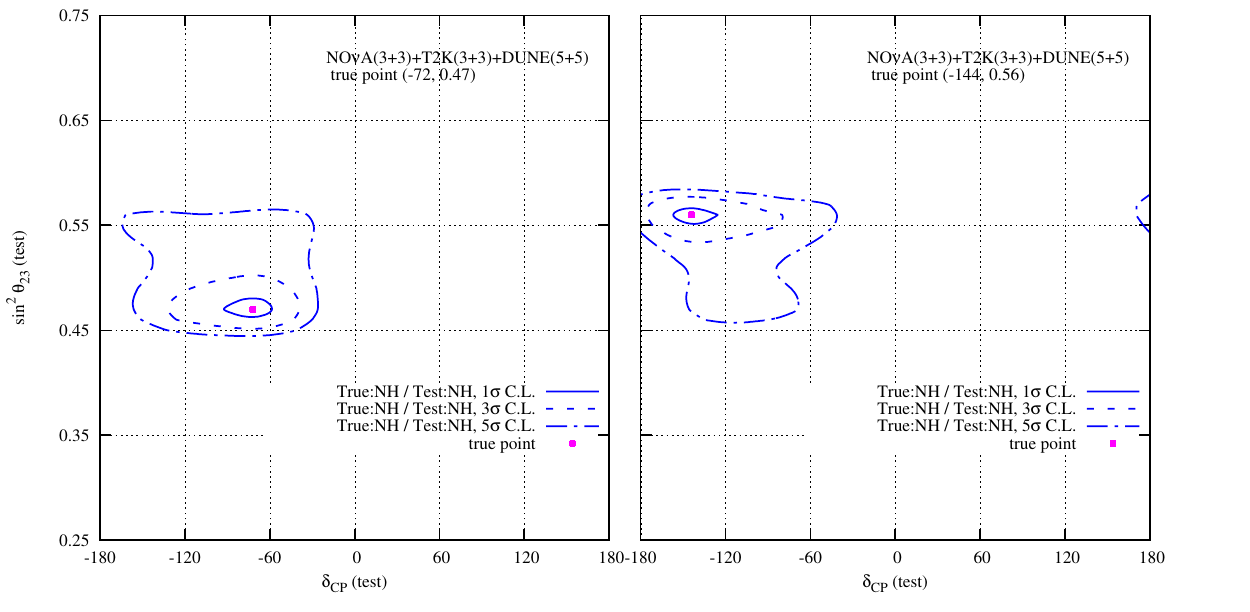}
\caption{\footnotesize{Regions of $\sin^2\theta_{23}$ vs. $\delta_{cp}$ parameter space. The left (right) column represents for LO (HO) for NO$\nu$A $(3\nu+3\bar{\nu})$ + 
T2K $(3\nu+3\bar{\nu})$ + DUNE $(5\nu+5\bar{\nu})$.}}
\label{dune5+5}
\end{figure}

We have also done a simulation of DUNE  $(5\nu+5\bar{\nu})$ run to check how well 
CP violation can be established. The results are shown in 
figure~\ref{dune5+5}. The left panel is for the LO solution and the right panel 
is for the HO solution. These figures are the result
of \nova $(3\nu+3\bar{\nu})$, T2K $(3\nu+3\bar{\nu})$ and
DUNE $(5\nu+5\bar{\nu})$ runs. The plots in these figures denote $1~\sigma$, $3~\sigma$
and $5~\sigma$ allowed contours. We note that, for LO solution, CP-violation can be established at $5~\sigma$. But, for the HO solution, $\dcp= 180^\circ$ is not ruled out at $5~\sigma$. The addition of $(5\nu+5\bar{\nu})$ run of DUNE
also helps in distinguishing between the two solutions at $3~\sigma$ level.   

This result can be understood from the point of view of changes
in $\nu_e$ appearance events induced by matter effects, octant 
effects and $\dcp$ effects on the reference point of vacuum oscillations,
maximal mixing and no CP-violation. \nova experiment observes 
certain number of $\nu_e$ appearance events. This number is moderately larger
than the number of such events expected from the reference point. The question is: which of the three effects
is contributing to this excess? The analysis of the data prefers 
NH as the hierarchy but has no preference for either octant. Since
the excess number of $\nu_e$ events is fixed, we can consider two
possibilities:
\begin{itemize}
\item
Part of the excess is because $\theta_{23}$ is in the higher octant.
That means only a limited part of the excess is due to $\dcp$ effect.
Values of $\dcp$ in the lower half plane lead to an increase in $\nu_e$
appearance, with the maximum increase coming for the case of $\dcp = -90^\circ$. 
If the $\dcp$ effects are to lead to only a modest excess, then
the preferred value of $\dcp$ will be away from $-90^\circ$. In the 
present case it is $-144^\circ$, closer to the CP conserving value of
$-180^\circ$ than the maximal CP-violating value of $-90^\circ$.
\item
If $\theta_{23}$ is in the lower octant, then the octant effects
{\bf suppress} $\nu_e$ events. To obtain the observed excess, then
the $\dcp$ effects have to compensate this suppression and provide
the excess. In such a situation, the preferred value of $\dcp$ will
be in the neighbourhood of maximal CP-violation. In the present case, 
it is $-72^\circ$, closer to maximal CP-violation rather than the CP conserving
value of $0/180^\circ$.
\end{itemize} 
It is, of course, easier to establish CP-violation if the 
value of $\dcp$ is close to $-90^\circ$.

\begin{table}
\begin{center}
\begin{tabular}{ |c|c|c|c| }
\hline
\hline
Hierarchy--$\sin^2\theta_{23}$--$\delta_{cp}$ & Signal eve. & Bg eve. & Total eve. \\
\hline\hline      
NH--0.56-- -144 & 379.72 & 97.57 & 477.29 \\
\hline   
NH--0.47-- -72 & 350.14 & 97.63 & 447.77 \\
\hline   
\hline
\end{tabular}
\caption{\footnotesize {Number of expected $\nu_e$ appearance events for one year $\nu$ run of DUNE, for the two solutions in ref.~\cite{Jan-2018,NOvA:2018gge}.}}
\label{DUNE-Nu-table-2}
\end{center}
\end{table}

\begin{table}
\begin{center}
\begin{tabular}{ |c|c|c|c| }
\hline\hline   
Hierarchy--$\sin^2\theta_{23}$--$\delta_{cp}$ & Signal eve. & Bg eve. & Total eve. \\
 \hline\hline      
NH--0.56-- -144 & 60.36 & 57.00 & 117.36 \\
\hline   
NH--0.47-- -72 & 49.71 & 57.03 & 106.74 \\
\hline   
\hline
\end{tabular}
\caption{\footnotesize {Number of expected $\bar{\nu}_e$ appearance events for one year $\bar{\nu}$ run of DUNE, for the two solutions in ref.~\cite{Jan-2018,NOvA:2018gge}.}}
\label{DUNE-barNu-table-2}
\end{center}
\end{table}

\section{Conclusions}

In this report, we have discussed the degeneracies present in the
\nova data. Before the measurement of $\theta_{13}$, $\pme$ had an eight fold degeneracy, 
arising due to the three unknown binomial variables: hierarchy, octant of $\theta_{23}$ and half plane of $\dcp$. However, 
the precision measurement of $\theta_{13}$ has split this degeneracy into 
the pattern $(1+3+3+1)$. If the observed number of $\nu_e$ events 
are well above or well below those expected from the reference point
of vacuum oscillations with maximal $\theta_{23}$ and no CP-violation,
then one can uniquely determine the hierarchy, octant and $\dcp$. If the difference between observed events and those expected 
from reference point is moderate, then, in general, there will be three degenerate solutions: One solution whose octant is distinct from that of the other two,
a second solution whose half plane of $\dcp$ is distinct from that of other two and a third solution whose hierarchy is distinct from that of the other two.
The early neutrino data of NO$\nu$A \cite{Adamson:2017gxd}, which showed a modest excess of $\nu_e$ appearance events,
gave rise to the three solutions,
(NH, LO, $-90^\circ$), (NH, HO, $135^\circ$) and (IH, HO, $-90^\circ$). These solutions do indeed have the pattern described above. 

In this report, we have shown that \nova will not be able to make a distinction between any of these three solutions.
The two higher octant solutions are completely degenerate with respect
to both neutrino and anti-neutrino data of NO$\nu$A. The lower octant
solution is distinct from the point of view of anti-neutrino data but 
the expected $\bar{\nu}_e$ events are quite small. The corresponding large
statistical errors prevent a clean isolation of this solution.
However, the addition of one year of neutrino data from DUNE can effectively isolate each of these 
three solutions at $3~\sigma$. 
The two NH solutions can be discriminated from the IH solution because of the large matter effects in DUNE.
Between the two NH solutions of different octants, 
both the anti-neutrino data of \nova and the neutrino data of DUNE have a
moderate discriminating capability. 
The synergy between these two sets of data is capable of providing a $3~\sigma$ discrimination 
between these two NH solutions.

Later data of NO$\nu$A, based on a more refined signal identification
algorithm \cite{Jan-2018,NOvA:2018gge}, has only two degenerate solutions: both with NH but with
different octants, where $\sin^2 \tz$ values in both cases are closer to maximal mixing.
This is a consequence of the new procedure, which has 
identified a larger number of signal events leading to a fairly large
excess of $\nu_e$ events compared to the expectation from the reference
point. A significant part of this excess occurs due to the matter effects
of NH. There are two possibilities to explain the remainder of the excess:
\begin{itemize}
\item
part of it is due to higher octant value of $\theta_{23}$ and part of it
is due to the $\dcp$ in lower half plane but well away from the maximal 
CP-violation of $-90^\circ$ 
\item 
a small suppression due to lower octant value of $\theta_{23}$ and a 
moderately large increase due to $\dcp$ being in the neighbourhood of
the maximal CP-violation value $-90^\circ$.
\end{itemize}
We found that neither \nova nor \nova + T2K is capable of distinguishing between
these two solutions at $3~\sigma$ level nor can they rule out the wrong hierarchy.
But addition of one year of neutrino data of DUNE is capable of ruling out the 
wrong hierarchy at $3~\sigma$ level but is unable to provide a similar discrimination
between the two solutions. Addition of a $(5\nu+5\bar{\nu})$ run of the DUNE experiment 
can distinguish between the solutions at $3~\sigma$. It can also establish CP-violation 
at $5~\sigma$ level for the lower octant solution but not for the higher octant solution. 
This occurs because the $\dcp$ value of the lower octant solution is closer to maximal CP-violation.

\section*{Acknowledgements}
SP thanks S\~ao Paulo Research Foundation (FAPESP)
for the support through Funding Grants No. 2014/19164-6 and
No. 2017/02361-1. UR thanks Council for Scientific and Industrial Research (CSIR), Government of India and Industrial 
Research and Consultancy Center (IRCC), IIT Bombay for financial support.

\appendix 

\section{Effect of systematic uncertainties}
In this appendix, we discuss the handling of systematic errors in 
our calculations and their effect on the results. The software
GLoBES includes systematic errors in 
\begin{itemize}
\item signal events number,
\item background events number, 
\item energy reconstruction in signal events and
\item energy reconstruction in background events.
\end{itemize}
These errors are defined separately for the neutrino run and for the
anti-neutrino run. 

We performed the calculations of \nova data
with three different sets of systematic errors. In the first 
case we set all systematic errors to zero. In the second case,
we have taken the systematic
error in signal events number to be $5\%$ and that in background
events number to be $10\%$ for both $\nu$ and $\bar{\nu}$ runs
but did not consider systematic error in energy reconstruction at all. 
In the third case, we took the systematic error in energy reconstruction
to be equal to systematic error in the events number for both signal
and for background.  

We find that the allowed regions in $\sin^2 \tz-\dcp$ plane, in
the second and third cases are identical. That is, inclusion of
systematic errors in energy reconstruction has no effect on the 
allowed parameter regions. The allowed region in the first case
is smaller by some $10\%$ or so, meaning that the inclusion
of systematic errors does have some effect on the determination of
parameters. We believe these results lead to the following conclusion.
The parameter regions in $\sin^2 \tz-\dcp$ plane are controlled 
by $\nu_e$ and $\bar{\nu}_e$ appearance data. Since the event 
numbers in both these channels are small (being proportional 
to $\sin^2 2 \theta_{13}$) the errors are dominated by statistics
and effect of systematic errors is small. In particular,  
the current parameter regions are essentially determined by the total number
of appearance events and the spectral information has only a 
secondary role. Hence the systematic error in energy reconstruction
has negligible effect on the allowed parameter region.

\bibliographystyle{JHEP}
\bibliography{referenceslist}
\end{document}